\documentstyle[amstex,prl,aps]{revtex}

\tightenlines

\begin{document}

\begin{center}
{\LARGE Spin 1 fields in Riemann-Cartan space-times $via$
Duffin-Kemmer-Petiau theory\vspace{1cm}\\[0pt]
}{\large R. Casana}$^{a}${\large ,\ J. T. Lunardi}$^{b}${\large ,\ B. M.
Pimentel}$^{a}${\large \ and R. G. Teixeira}$^{c}${\Large \vspace{1cm}}%
{\large \\[0pt]
}$^{a}${\normalsize Instituto de F\'{\i}sica Te\'{o}rica, Universidade
Estadual Paulista\\[0pt]
Rua Pamplona 145, 01405-900, S\~{a}o Paulo, SP, Brazil\bigskip \\[0pt]
}$^{b}${\normalsize Departamento de Matem\'{a}tica e Estat\'{\i}stica,
Universidade Estadual de Ponta Grossa\\[0pt]
Av. Gal. Carlos Cavalcanti 4748, 84032-900, Ponta Grossa, PR, Brazil\bigskip 
\\[0pt]
}$^{c}${\normalsize Faculdade de Tecnologia e Ci\^{e}ncias Exatas,
Universidade S\~{a}o Judas Tadeu\\[0pt]
Rua Taquari 546, 03166-000, S\~{a}o Paulo, SP, Brazil\\[0pt]
}
\end{center}

\medskip 
\abstract{We consider massive spin 1 fields, in Riemann-Cartan
space-times, described by Duffin-Kemmer-Petiau theory. We show that
this approach induces a coupling between the spin 1 field and
the space-time torsion which breaks the usual equivalence with the
Proca theory, but that such equivalence is preserved
in the context of the Teleparallel Equivalent of General
Relativity.} \bigskip \newline
\bigskip \newline
{\large Dedicated to Professor Gerhard Wilhelm Bund on the occasion of his
70th birthday.}\bigskip

\section{Introduction}

The Duffin-Kemmer-Petiau (DKP) equation is a first order relativistic
equation, similar to the Dirac's one, which describes fields of spin 0 and 1 
\cite{Duffin,Kemmer,Petiau,Umezawa}. The basic aspects and properties of the
DKP equation which are necessary to the comprehension of this work can be
found in the references \cite{Lunardi 1,Lunardi 2}, where it was adopted the
same metric signature. For a historical review covering the theory until the
decade of 1970 we refer to reference \cite{Nieto}.

Recently there have been a renewed interest in DKP theory. For instance, it
has been studied in the context of QCD \cite{Gribov}, covariant hamiltonian
dynamics \cite{Kanatchikov}, in the Causal Approach \cite{Lunardi 3}, in the
context of five-dimensional galilean covariance \cite{Esdras}, in the
scattering $K^{+}$-nucleus \cite{Prog}, and in curved space-times \cite
{Lunardi 1,Red'kov}, among other situations.

One important question concerning the DKP theory is about the equivalence or
not of its spin 0 and 1 sectors to the Klein-Gordon (KG) and the Proca
theories, respectively. This is an old question for which, nowadays, still
lacks a complete answer. Recently, there have been some efforts to give
strict proofs of equivalence between the KG equation and the spin 0 sector
of the DKP equation in various situations \cite{Fainberg 1,Fainberg
2,Fainberg 3}. In the same context, some aspects regarding the minimal
interaction with the electromagnetic field have been clarified \cite{Lunardi
2,Nowakowski}. Moreover, the equivalence between the DKP and the KG and the
Proca fields for spin 0 and 1 has also been proved in the context of a
riemannian space-time \cite{Lunardi 1}.

On the other hand, the study of the DKP theory for massive spin 0 fields
minimally coupled to Riemann-Cartan (RC) space-time has been carried out in
the reference \cite{Lunardi 4}, where it was shown that, in the context of
Einstein-Cartan theory, the DKP formalism naturally induces an interaction
between the spin 0 field and the space-time torsion, breaking the
equivalence with the KG equation, which does not present any interaction
with torsion. In the same reference it was also discussed the conceptual
differences between this kind of interaction and that which appears in the
context of the Teleparallel Equivalent of General Relativity
(Teleparallelism theory), where the spin 0 sector of DKP field and KG field
are completely equivalent.

Our aim in this paper is to complete this analysis by studying massive spin
1 fields in Riemann-Cartan space-times using the DKP theory, both in the
context of Einstein-Cartan and Teleparallelism theories, and comparing the
results with those obtained in the framework of Proca's field approach. In
the next section we present the DKP theory in Minkowski space-time. In
section 3 we introduce minimally coupling to the Riemann-Cartan space-time
and select the spin 1 sector in order to compare the results with those
obtained through Proca's field. In section 4 we analyse both the DKP and the
Proca fields in the context of the Teleparallelism theory and, in section 5,
we present our concluding remarks.

\section{DKP field in Minkowski space-times}

The Duffin-Kemmer-Petiau equation in Minkowski space ($M^{4}$) is given by 
\begin{equation}
i\beta ^{a}\partial _{a}\,\psi -m\,\psi =0,  \label{eq1x}
\end{equation}
where $a=0,1,2,3$ are spatiotemporal Minkowski indexes\footnote{%
The Latin alphabet will be used {\bf throughout} this paper to indicate
Minkowski indexes while Riemman-Cartan indexes will be indicated by Greek
letters.}. The matrices $\beta ^{a}$ obey the DKP algebra, given by 
\begin{equation}
\beta ^{a}\beta ^{b}\beta ^{c}+\beta ^{c}\beta ^{b}\beta ^{a}=\beta ^{a}\eta
^{bc}+\beta ^{c}\eta ^{ba};  \label{eq2x}
\end{equation}
with $\eta ^{ab}$ being the metric tensor of Minkowski space-time with
signature $\left( +---\right) $.

The DKP equation (\ref{eq1x}) is very similar to the Dirac's one, but the
algebraic properties of $\beta ^{a}$ matrices, which have no inverses, make
it more difficult to deal with. This equation can also be obtained from the
Lagrangian density 
\begin{equation}
{\cal L}=\frac{i}{2}\;\overline{\psi }\beta ^{a}\partial _{a}\psi -\frac{i}{2%
}\;\left( \partial _{a}\overline{\psi }\right) \beta ^{a}\psi -m\overline{%
\psi }\psi ,  \label{eq6x}
\end{equation}
where $\overline{\psi }=\psi ^{\dagger }\eta ^{0}$, $\eta ^{0}=2\left( \beta
^{0}\right) ^{2}-1$ and we choose $\beta ^{0}$\ to be hermitian and $\beta
^{i}$\ $\left( i=1,2,3\right) $\ anti-hermitian.

It can be shown \cite{Kemmer,Umezawa} that DKP algebra has only 3
inequivalent irreducible representations, with degrees 1, 5 and 10. The
first one is trivial $\left( \beta ^{a}=0\right) $, having no physical
significance, while the other two represent fields of spin 0 and 1,
respectively. Moreover, for any representation, one can define a set of
operators (Umezawa's ``projectors'') which selects the scalar and vector
sectors of the DKP field \cite{Umezawa}.

\section{The Einstein-Cartan theory}

From now on we follow the definitions and notations from the references \cite
{Lunardi 1,Lunardi 4} for the DKP field in curved manifolds. We remember
that the covariant derivative $\nabla $ in the Einstein-Cartan theory (which
assumes a Riemman-Cartan space-time geometry) has an affine connection $%
\Gamma _{\mu \nu }{}^{\alpha }$, not necessarily symmetric in the lower
indexes, whose antisymmetric part $Q_{\mu \nu }{}^{\alpha }$ is the Cartan
torsion, i. e., 
\begin{equation}
Q_{\mu \nu }{}^{\alpha }=\frac{1}{2}\left( {\Gamma }_{\mu \nu }{}^{\alpha }-{%
\Gamma }_{\nu \mu }{}^{\alpha }\right) .
\end{equation}

Then we can write the affine connection as 
\begin{equation}
{\Gamma }_{\mu \nu }{}^{\alpha }=\stackrel{r}{\Gamma }_{\mu \,\nu
}{}^{\alpha }-K_{\mu \nu }{}^{\alpha },  \label{eqxx1}
\end{equation}
where $\stackrel{r}{\Gamma }_{\mu \nu }{}^{\alpha }$ is the Christoffel
symbol (or the riemannian part of the connection) and $K_{\mu \nu
}{}^{\alpha }$ is the contorsion tensor, defined as 
\begin{equation}
K_{\mu \nu }{}^{\alpha }=-Q_{\mu \nu }{}^{\alpha }-Q^{\,\alpha }{}_{\nu \mu
}+Q_{\mu }{}^{\alpha }{}_{\nu }.  \label{eqxx6}
\end{equation}

The covariant derivative of the DKP field is given by \cite{Lunardi 1} 
\begin{equation}
{\nabla }_{\mu }\psi ={D}_{\mu }\psi =\left( \partial _{\mu }+\frac{1}{2}%
\omega _{\mu ab}S^{ab}\right) \psi  \label{eq1c}
\end{equation}
and is formally similar to that of Dirac's field \cite{Sabbata}. In this
expression $\omega _{\mu ab}$ is the spin connection and $S_{ab}=\beta
_{a}\beta _{b}-\beta _{b}\beta _{a}$. The matrices $\beta ^{\mu }$ are
defined through contraction with the tetrad (or vierbein) fields $e^{\mu
}{}_{a}$, i.e. $\beta ^{\mu }=e^{\mu }{}_{a}\beta ^{a}$, and they satisfy
the generalized DKP algebra 
\begin{equation}
\beta ^{\mu }\beta ^{\nu }\beta ^{\alpha }+\beta ^{\alpha }\beta ^{\nu
}\beta ^{\mu }=\beta ^{\mu }g^{\nu \alpha }+\beta ^{\alpha }g^{\nu \mu },
\label{eq1b}
\end{equation}
where $g^{\nu \mu }$ is the Riemann-Cartan metric tensor. In the
Einstein-Cartan theory the spin connection can be written in terms of the
affine connection and the tetrad field as \cite{Sabbata} 
\begin{equation}
\omega _{\mu }{}^{ab}=\gamma _{\mu }{}^{ab}-K_{\mu }{}^{ba},  \label{eq33c}
\end{equation}
where 
\begin{equation}
K_{\mu }{}^{ba}=-K_{\mu }{}^{ab}=e^{\alpha a}e^{\beta b}K_{\mu \alpha \beta
},  \label{eq33d}
\end{equation}
while $\gamma _{\mu }{}^{ab}$ is the riemannian part of the spin connection,
given by 
\begin{equation}
\gamma _{\mu }{}^{ab}=-\gamma _{\mu }{}^{ba}=e_{\mu i}\left(
C^{abi}-C^{bia}-C^{iab}\right) ,  \label{eq33f}
\end{equation}
being $C^{abi}$ the Ricci rotation coefficients\footnote{%
The brackets in this expression denote {\bf antisymmetrization} of the
enclosed indexes.} 
\begin{equation}
C_{ab}{}^{i}=e^{\mu }{}_{a}\left( x\right) e^{\nu }{}_{b}\left( x\right)
\partial _{\lbrack \mu }{}e_{\nu ]}{}^{i}\,.  \label{eq33g}
\end{equation}

The Lagrangian density for the DKP field minimally coupled \cite{misner,hehl}
to the Riemann-Cartan manifold is given by \cite{Lunardi 4} 
\begin{equation}
{\cal L}=\sqrt{-g}\left[ \frac{i}{2}\;\overline{\psi }\beta ^{\mu }{\nabla }%
_{\mu }\psi -\frac{i}{2}\;\left( {\nabla }_{\mu }\overline{\psi }\right)
\beta ^{\mu }\psi -m\overline{\psi }\psi \right] ,  \label{eq1}
\end{equation}
from which we get the equation of motion for the massive DKP field in the
Einstein-Cartan theory as 
\begin{equation}
i\,\beta ^{\mu }{\nabla }_{\mu }\psi +\frac{i}{2}K_{\sigma \mu }{}^{\sigma
}\beta ^{\mu }\psi -m\psi =0.  \label{eq3}
\end{equation}
We can promptly see that this equation differs from the one that would be
obtained from the Minkowskian DKP equation of motion (\ref{eq1x}) through
the minimal coupling procedure as is usual in Einstein-Cartan theory \cite
{Saa,Saa2}.

\subsection{Spin $1$ sector}

Now we use the Umezawa's ``projectors'' $R^{\mu }$ and $R^{\mu \nu }$ in
order to analyse the spin $1$ sector of the theory. We remember that each
component of $R^{\mu }\psi $ is a vector and each one of $R^{\mu \nu }\psi $
is a second rank antisymmetric tensor \cite{Umezawa,Lunardi 1,Lunardi 2}.
Applying these operators on the left of equation of motion (\ref{eq3}) we
get, respectively, 
\begin{mathletters}
\label{eq7}
\begin{eqnarray}
m\,R^{\mu }\psi =i\,\tilde{{\cal D}}_{\nu }\,(R^{\mu \nu }\psi ), \\
m\,(R^{\mu \nu }\psi )=i\left( \tilde{{\cal D}}^{\nu }R^{\mu }\psi -\tilde{%
{\cal D}}^{\mu }R^{\nu }\psi \right) ,
\end{eqnarray}
where the derivative operator $\tilde{{\cal D}}_{\nu }$ is defined as 
\end{mathletters}
\begin{equation}
\tilde{{\cal D}}_{\mu }={\nabla }_{\mu }+\frac{1}{2}\;K_{\sigma \mu
}{}^{\sigma }.
\end{equation}

Combining both the equations (\ref{eq7}) we get the equation of motion for
the massive vector field $R^{\mu }\psi $\ 
\begin{equation}
\tilde{{\cal D}}_{\beta }\tilde{{\cal D}}_{\alpha }T^{\alpha \beta \mu
}+m^{2}\,(R^{\mu }\psi )=0,  \label{eq10a}
\end{equation}
or, written explicitly, 
\begin{equation}
\left( {\nabla }_{\beta }+\frac{1}{2}\;K_{\sigma \beta }{}^{\sigma }\right)
\left( {\nabla }_{\alpha }+\frac{1}{2}\;K_{\sigma \alpha }{}^{\sigma
}\right) T^{\alpha \beta \mu }+m^{2}\,(R^{\mu }\psi )=0,  \label{eq10b}
\end{equation}
where we have defined $T^{\alpha \beta \mu }=g^{\alpha \beta }(R^{\mu }\psi
)-g^{\alpha \mu }(R^{\beta }\psi )$.

\subsubsection{Proca's field}

In Minkowski space-time the Lagrangian density for Proca's field is given by 
\begin{equation}
{\cal L}_{_{\!{\cal M}}}=-\frac{1}{4}F_{ab}F^{ab}+\frac{1}{2}%
\,m^{2}A_{a}A^{a},  \label{eq14}
\end{equation}
where $F_{ab}=\partial _{a}A_{b}-\partial _{b}A_{a}$ is the field strength
tensor.

When the procedure of minimal coupling to the Riemann-Cartan manifold is
performed on the above Lagrangian we obtain 
\begin{equation}
{\cal L}_{_{\!{\cal U}}}=\sqrt{-g}\left( -\frac{1}{4}\;F_{\mu \nu }F^{\mu
\nu }+\frac{1}{2}\;m^{2}A_{\mu }A^{\mu }+F_{\mu \nu }Q_{\quad \!\sigma
}^{\mu \nu }A^{\sigma }-Q_{\mu \nu }^{\quad \!\rho }Q_{\quad \!\sigma }^{\mu
\nu }A_{\rho }A^{\sigma }\right) ,  \label{eq15}
\end{equation}
where 
\begin{equation}
F_{\mu \nu }=\partial _{\mu }A_{\nu }-\partial _{\nu }A_{\mu }.
\label{eq15v}
\end{equation}
The equations of motion for the field $A_{\mu }$, obtained from this
Lagrangian, are given by 
\begin{equation}
\left( {\nabla }_{\nu }+K_{\sigma \nu }{}^{\sigma }\right) \left( {\nabla }%
^{\nu }A^{\mu }-\nabla ^{\mu }A^{\nu }\right) +m^{2}A^{\mu }=0,
\label{eq11a}
\end{equation}
or 
\begin{equation}
\left( {\nabla }_{\beta }+K_{\sigma \beta }{}^{\sigma }\right) {\nabla }%
_{\alpha }T^{\alpha \beta \mu }+m^{2}A^{\mu }=0,  \label{eq11b}
\end{equation}
being $T^{\alpha \beta \mu }=g^{\alpha \beta }A^{\mu }-g^{\alpha \mu
}A^{\beta }$. These equations show the usual interaction between Proca%
\'{}%
s field and torsion, as known in literature, and are very different from
equations (\ref{eq10a}) and (\ref{eq10b}) given by the DKP approach for the
spin $1$\ field in the Einstein-Cartan theory.

\subsection{Comparison between DKP and Proca fields}

In order to compare the results obtained through DKP field with those from
Proca's one we will use the explicit representation of $\beta $ matrices
given in \cite{Lunardi 2}. With this representation we have 
\[
R^{\mu }\psi =\left( 
\begin{array}{c}
\psi ^{\mu } \\ 
0_{9x1}
\end{array}
\right) \quad ,\quad R^{\mu \nu }\psi =\left( 
\begin{array}{c}
\psi ^{\mu \nu } \\ 
0_{9x1}
\end{array}
\right) \quad ,\quad \mu =0,1,2,3, 
\]
where the ten-component DKP field is 
\[
\psi =\left( \psi ^{0},\psi ^{1},\psi ^{2},\psi ^{3},\psi ^{4},\psi
^{5},\psi ^{6},\psi ^{7},\psi ^{8},\psi ^{9}\right) ^{T}. 
\]
We have denoted the field components as 
\[
\psi ^{4}=\psi ^{23},\;\psi ^{5}=\psi ^{31},\;\psi ^{6}=\psi ^{12},\;\psi
^{7}=\psi ^{10},\;\psi ^{8}=\psi ^{20},\;\text{and}\;\psi ^{9}=\psi ^{30}, 
\]
so that we get from equation (\ref{eq7}) 
\begin{equation}
\psi ^{\mu \nu }=\frac{i}{m}\left( \tilde{{\cal D}}^{\nu }\psi ^{\mu }-%
\tilde{{\cal D}}^{\mu }\psi ^{\nu }\right) .
\end{equation}

Finally, we can define $\psi ^{\mu }=\sqrt{\frac{m}{2}}\,A^{\mu }$, with $%
A^{\mu }$ being a real vector field. Then, using the explicit form of $\psi $
given above, the DKP Lagrangian (\ref{eq1}) can be rewritten in terms of the
fields $A^{\mu }$ as 
\begin{equation}
{\cal L}=\sqrt{-g}\left( -\frac{1}{4}\;U_{\mu \nu }U^{\mu \nu }+\frac{1}{2}%
\;m^{2}A_{\mu }A^{\mu }\right) ,
\end{equation}
where 
\begin{equation}
U^{\mu \nu }=\tilde{{\cal D}}^{\mu }A^{\nu }-\tilde{{\cal D}}^{\nu }A^{\mu
}\,.  \label{nonmin}
\end{equation}
The Lagrangian above can be written explicitly as 
\begin{eqnarray}
{\cal L} &=&\sqrt{-g}\,\left( -\frac{1}{4}\;F_{\mu \nu }F^{\mu \nu }+\frac{1%
}{2}\;m^{2}A_{\mu }A^{\mu }+F_{\mu \nu }Q_{\quad \!\sigma }^{\mu \nu
}A^{\sigma }+\right.  \nonumber \\
&&\left. -Q_{\mu \nu }^{\quad \!\rho }Q_{\quad \!\sigma }^{\mu \nu }A_{\rho
}A^{\sigma }-\frac{1}{4}\;F_{\mu \nu }\Sigma ^{\mu \nu }+\frac{1}{2}\;\Sigma
_{\mu \nu }Q_{\quad \!\sigma }^{\mu \nu }A^{\sigma }-\frac{1}{16}\;\Sigma
_{\mu \nu }\Sigma ^{\mu \nu }\right) ,  \label{eq26}
\end{eqnarray}
where the tensor $\Sigma ^{\mu \nu }$ is defined as 
\begin{equation}
\Sigma _{\mu \nu }=K_{\sigma \mu }^{\quad \!\sigma }A_{\nu }-K_{\sigma \nu
}^{\quad \!\sigma }A_{\mu }\,.
\end{equation}

Due to the presence of the last three terms in the above Lagrangian it is
not equivalent to the minimally coupled Proca's Lagrangian (\ref{eq15}). In
fact, it is straightforward to see from (\ref{nonmin}) that the minimally
coupled DKP Lagrangian (equation (\ref{eq1}) or (\ref{eq26})) is equivalent
to that obtained from the Minkowskian Proca Lagrangian by means of the
following non-minimal substitution\footnote{%
Besides, of course, the global multiplication by the factor $\sqrt{-g}$.} 
\begin{equation}
\partial _{a}\rightarrow \tilde{{\cal D}}_{\mu }.  \label{nms}
\end{equation}

\section{DKP field in the Teleparallel theory}

The analysis of DKP field coupled to gravitation in the framework of the
Teleparallel Equivalent of General Relativity \cite{hayashi} was developed
in reference \cite{Lunardi 4}. Specifically, it was considered the spin 0
sector of the theory and the results were compared to those obtained from KG
field. Here we will extended such analysis to the spin 1 sector of the DKP
theory and compare the results with those obtained from Proca's field.

We remember that the DKP Lagrangian minimally coupled to the Riemann
space-time (which is the space-time of General Relativity and is a special
case of the Riemann-Cartan space-time whose torsion vanishes identically)
can be equivalently written in terms of the Teleparalell structure, which
describes fields in a Weitzenb\"{o}ck space-time (another special case of a
Riemann-Cartan space-time whose curvature vanishes identically).

To construct the equations for DKP field in the Teleparallel framework we
will start from the corresponding equations for DKP theory in General
Relativity, as given in reference \cite{Lunardi 2}. Then, we will simply
replace the riemannian quantities in these equations by the corresponding
teleparallel ones, according to the rules \cite{Lunardi 4} 
\begin{mathletters}
\label{eq70}
\begin{eqnarray}
\stackrel{_{r}}{\Gamma }_{\alpha \beta }{}^{\mu }\rightarrow {\Gamma }%
_{\alpha \beta }{}^{\mu }+K_{\alpha \beta }{}^{\mu }\,, \\
\stackrel{_{r}}{\nabla }_{\mu }\psi \rightarrow \left( \partial _{\mu }-%
\frac{1}{2}K_{\mu \alpha \beta }S^{\alpha \beta }\right) \psi , \\
\stackrel{r}{\nabla }_{\mu }\overline{\psi }\rightarrow \overline{\psi }%
\left( \overleftarrow{\partial }_{\mu }+\frac{1}{2}K_{\mu \alpha \beta
}S^{\alpha \beta }\right) \,.
\end{eqnarray}

Making so, the General Relativity Lagrangian written in terms of the
Teleparalell structure is given by \cite{Lunardi 4} 
\end{mathletters}
\begin{equation}
{\cal L}=e\left\{ \frac{i}{2}\left[ \overline{\psi }\beta ^{\mu }\left(
\partial _{\mu }-\frac{1}{2}K_{\mu \alpha \beta }S^{\alpha \beta }\right)
\psi -\overline{\psi }\left( \overleftarrow{\partial }_{\mu }+\frac{1}{2}%
K_{\mu \alpha \beta }S^{\alpha \beta }\right) \beta ^{\mu }\psi \right] -m%
\overline{\psi }\psi \right\} ,  \label{eq72}
\end{equation}
from which we get the equation of motion 
\begin{equation}
i\beta ^{\mu }\left( \partial _{\mu }-\frac{1}{2}K_{\mu \alpha \beta
}S^{\alpha \beta }\right) \psi -m\psi =0\,.  \label{eq76}
\end{equation}

\subsection{Spin 1 sector}

Now we apply the operators $R^{\mu }$ and $R^{\mu \nu }$ on the equation of
motion (\ref{eq76}). From the results of \cite{Lunardi 2} and making use of
the rules (\ref{eq70}) we get 
\begin{mathletters}
\label{eqB}
\begin{eqnarray}
R^{\mu }\psi =\frac{i}{m}{\cal D}_{\nu }(R^{\mu \nu }\psi )  \label{eqB-a} \\
R^{\mu \nu }\psi =\frac{i}{m}\left[ {\cal D}^{\mu }(R^{\nu }\psi )-{\cal D}%
^{\nu }(R^{\mu }\psi )\right] \,,  \label{eqB-b}
\end{eqnarray}
where the covariant derivative $D_{\mu }$ is defined as\footnote{%
The derivative ${\cal D}_{\mu }$ is sometimes referred to as the {\it %
teleparallel version} of General Relativity's covariant derivative \cite
{Andrade 2}, because it is nothing more than the General Relativity
covariant derivative written in terms of the teleparallel quantities.} 
\end{mathletters}
\begin{equation}
{\cal D}_{\mu }=\nabla _{\mu }+K_{\mu }\,,\qquad \text{with}\qquad \nabla
_{\mu }=\partial _{\mu }+\Gamma _{\mu }\,,
\end{equation}
In these last expressions $\Gamma _{\mu }$ is the Cartan connection \cite
{Lunardi 4} 
\[
\Gamma _{\mu \nu }{}^{\alpha }=e^{\alpha }{}_{i}\partial _{\mu }e_{\nu
}{}^{i}, 
\]
which is associated with the Weitzenb\"{o}ck space, and $K$ is the
corresponding contorsion tensor, as given by equation (\ref{eqxx6}).

Combining the equations (\ref{eqB-a}) and (\ref{eqB-b}) we obtain the
equation of motion for the spin 1 DKP field in the Teleparallel framework 
\begin{equation}
{\cal D}_{\nu }\left[ {\cal D}^{\nu }(R^{\mu }\psi )-{\cal D}^{\mu }(R^{\nu
}\psi )\right] +m^{2}(R^{\mu }\psi )=0\,.  \label{eqxx8}
\end{equation}

\subsection{Proca's field}

The General Relativity Lagrangian density of Proca's field, written in terms
of the Teleparallel structure, is given by 
\begin{equation}
{\cal L}=e\left( -\frac{1}{4}\;F_{\mu \nu }F^{\mu \nu }+\frac{1}{2}%
\;m^{2}A_{\mu }A^{\mu }\right) \,.  \label{eqxx9}
\end{equation}
This expression could be obtained from the Proca's Lagrangian in Minkowski
space by means of the following prescription \cite{Andrade 2} \footnote{%
Besides, we must multiply the whole lagrangian by a factor $e={\rm det}%
|e^{a}{}_{\mu }|=\sqrt{-g}$ to make it a scalar density.} 
\[
\partial _{a}\rightarrow {\cal D}_{\mu }\equiv \partial _{\mu }+\Gamma _{\mu
}+K_{\mu }=\nabla _{\mu }+K_{\mu }\,. 
\]
The stress tensor for the Proca's field in the Teleparallel structure is
given by 
\[
F_{\mu \nu }={\cal D}_{\mu }A_{\nu }-{\cal D}_{\nu }A_{\mu }\,. 
\]
By using the explicit form of $D_{\mu }$ and the definition of torsion and
contorsion tensors, it is an easy task to verify that 
\[
F_{\mu \nu }=\partial _{\mu }A_{\nu }-\partial _{\nu }A_{\mu }, 
\]
and from the Lagrangian (\ref{eqxx9}) we get the teleparallel version of the
Proca's equation 
\begin{equation}
{\cal D}_{\nu }F^{\nu \mu }+m^{2}A^{\mu }=0.  \label{eqxx12}
\end{equation}

Thus, comparing equations (\ref{eqxx8}) and (\ref{eqxx12}), we conclude that
the DKP and the Proca theories for massive spin 1 fields in the context of
the Teleparallel description of General Relativity are completely
equivalent. This is an expected result because both Proca theory and the
spin 1 sector of DKP theory give the same results in the context of General
Relativity, of which the Teleparallelism is an equivalent description.

\section{Concluding Remarks}

In the reference \cite{Lunardi 4} it was shown that the spin 0 sector of DKP
theory is not equivalent to Klein-Gordon theory in the context of
Einstein-Cartan theory with minimal coupling procedure. Differently to what
happens with the KG theory, in which does not appear any interaction between
the scalar field and the space-time torsion, in the DKP theory this
interaction naturally arises. It is interesting to notice that the concept
of a scalar field interacting naturally (i.e. through a minimal coupling
procedure) with torsion \cite{Lunardi 5} is useful in the context of a
quantum theory of matter fields in a Riemann-Cartan space-time because it
gives the possibility of constructing a renormalizable theory \cite
{shapiro,shapiro2}.

Here we completed this analysis by studying the massive spin 1 sector of DKP
theory. We showed that the spin 1 sector of DKP theory and Proca theory are
inequivalent in the context of Einstein-Cartan theory with minimal coupling.
Although in this context Proca's formalism allows an interaction between
massive spin 1 fields and the space-time torsion (see the Lagrangian (\ref
{eq15})), the DKP formalism presents a more general interaction with the
torsion, containing all the terms present in the Proca Lagrangian plus three
additional terms (see equation (\ref{eq26})).

Still extending the analysis of reference \cite{Lunardi 4}, we considered
the massive spin 1 sector of DKP theory in the framework of the Teleparallel
Equivalent of General Relativity. Nevertheless the fact that in this
framework the gravitational field is associated with the space-time torsion
and not with curvature, the DKP and the Proca approaches give identical
results. This was an expected result since both formalisms are equivalent in
the context of General Relativity.

Finally, as it is well known, the application of the minimal coupling
procedure to the Maxwell Lagrangian induces a coupling to the space-time
torsion which breaks the gauge invariance of the theory, a consequence which
is usually avoided by the introduction of non minimal couplings. From the
results of reference \cite{Lunardi 4} and those of the present work we saw
that, in the context of Einstein-Cartan theory, the DKP field with minimal
coupling is equivalent to performing non-minimal couplings in the KG or
Proca's fields. Then, it seems interesting to investigate if the use of DKP
field in the study of massless spin 1 fields\footnote{%
We observe that DKP formalism for massless spin 1 fields cannot be obtained
as a limiting case of the massive theory, as it happens in the case of
Maxwell and Proca theories.} on Einstein-Cartan backgrounds can give further
insights on the incompatibility between gauge invariance and the interaction
with torsion. This question is presently under our investigation.

\section{Acknowledgements}

B.M.P. and R.C. thank to CNPq for partial and full support, respectively.
J.T.L. and R.G.T. thank to Instituto de F\'{\i}sica Te\'{o}rica (IFT) for
granting them access to its facilities during the development of this work.

\end{document}